\documentclass[journal]{IEEEtran}

\usepackage{cite,bm,tikz,amssymb,amsthm,amsmath,graphicx,pgfplots,multirow,epstopdf,algorithm,algorithmicx,algpseudocode,url}
\usepackage[acronym]{glossaries}
\usetikzlibrary{arrows.meta,decorations.pathreplacing,plotmarks,shapes,arrows,positioning}
\pgfplotsset{major grid style={solid,white!90!black}}
\pgfplotsset{tick style={solid,white!0!black}}
\pgfplotsset{minor tick length=0.07cm}
\pgfplotsset{every tick label/.append style={font=\scriptsize}}
\pgfplotsset{tickwidth=0.07cm}

\let\originalleft\left
\let\originalright\right
\renewcommand{\left}{\mathopen{}\mathclose\bgroup\originalleft}
\renewcommand{\right}{\aftergroup\egroup\originalright}

\newcommand{\U}[1]{U\textsubscript{#1}}
\renewcommand{\S}[1]{S\textsubscript{#1}}

\pgfplotsset{/pgfplots/ybar legend/.style={
		/pgfplots/legend image code/.code={%
			\draw[##1,/tikz/.cd,yshift=-0.25em,rounded corners=0]
			(0cm,0cm) rectangle (8pt,0.8em);},
	},
}

\definecolor{mycolor1}{rgb}{0.121568627450980,0.466666666666667,0.705882352941177}
\definecolor{mycolor2}{rgb}{0.894117647058824,0.101960784313725,0.109803921568627}
\definecolor{mycolor3}{rgb}{1.000000000000000,0.498039215686275,0.054901960784314}
\definecolor{mycolor4}{rgb}{0.172549019607843,0.627450980392157,0.172549019607843}
\definecolor{mycolor5}{rgb}{0.666666666666667,0.200000000000000,0.607843137254902}
\definecolor{mycolor6}{rgb}{0.35,0.35,0.35}

\makeatletter \newcommand{\pgfplotsdrawaxis}{\pgfplots@draw@axis} \makeatother

\pgfplotsset{axis line on top/.style={
		axis line style=transparent,
		ticklabel style=transparent,
		tick style=transparent,
		axis on top=false,
		after end axis/.append code={
			\pgfplotsset{axis line style=opaque,
				ticklabel style=opaque,
				tick style=opaque,
				grid=none}
			\pgfplotsdrawaxis}
	}
}

\DeclareMathOperator*{\diag}{diag}
\DeclareMathOperator*{\tr}{tr}

\usepgfplotslibrary{external} 
\begin{document}

\title{Pilot Distributions for Joint-Channel Carrier-Phase Estimation in Multichannel Optical Communications}

\author{
	Arni~F.~Alfredsson,~\IEEEmembership{Student~Member,~IEEE},~Erik~Agrell,~\IEEEmembership{Fellow,~IEEE},\\Magnus~Karlsson,~\IEEEmembership{Senior~Member,~IEEE,~Fellow,~OSA},~and~Henk~Wymeersch,~\IEEEmembership{Member,~IEEE}%
\thanks{
A. F. Alfredsson, E. Agrell, and H. Wymeersch are with the Department of Electrical Engineering, Chalmers University of Technology, SE-41296 G\"{o}teborg, Sweden (e-mail: arnia@chalmers.se).

M. Karlsson is with the Photonics Laboratory, Department of Microtechnology and Nanoscience, Chalmers University of Technology, SE-41296 G\"{o}teborg, Sweden.

This work was supported by the Swedish Research Council (VR) via Grants 2014-6138 and 2018-03701 and the Knut and Alice Wallenberg Foundation via Grant 2018.0090.
}}
\maketitle

\begin{abstract}
	\boldmath
	Joint-channel carrier-phase estimation can improve the performance of multichannel optical communication systems. In the case of pilot-aided estimation, the pilots are distributed over a two-dimensional channel--time symbol block that is transmitted through multiple channels. However, suboptimal pilot distributions reduce the effectiveness of the carrier-phase estimation and thus result in unnecessary pilot overhead, which reduces the overall information rate of the system. It is shown that placing pilots identically in all channels is suboptimal in general. By instead optimizing the pilot distribution, the mean squared error of the phase-noise estimates can be decreased by over 90\% in some cases. Moreover, it is shown that the achievable information rate can be increased by up to 0.05, 0.16, and 0.41 bits per complex symbol for dual-polarization 20 GBd transmission of 64-ary, 256-ary, and 1024-ary quadrature amplitude modulation over 20 four-dimensional channels, respectively, assuming a total laser linewidth of 200 kHz.
\end{abstract}


\section{Introduction}
\label{sec:intro}

Laser phase noise is one of the main transmission impairments in fiber-optic communications. With the increased use of higher-order modulation formats \cite{8346089,6886985,Olsson:18,2019:chen}, which are inherently more sensitive to phase noise, it becomes crucial to estimate and compensate for this impairment effectively.
Traditionally, carrier-phase estimation (CPE) has been implemented using blind methods \cite{5447711}, operating without the aid of pilot symbols that are known to the receiver \cite{4814758,4298982}.
More recently, as fiber-optic systems have started being pushed to their limits in order to increase spectral efficiency and transmission reach, pilot-aided CPE methods \cite{Mazur:19,8859308,8695027,7384692,7301998} have become a popular option. This is due to their modulation-format transparency and their ability to operate at lower signal-to-noise ratios (SNRs) than blind methods \cite{Mazur:19}.

In addition to the use of higher-order modulation formats, coherent multichannel transmission has been a key enabler for spectrally efficient systems. Multichannel transmission has traditionally been implemented through wavelength-division multiplexing (WDM) and more recently also through space-division multiplexing (SDM). In such systems, lasers can be shared by multiple channels, e.g., in frequency-comb based WDM systems \cite{8327487} and in SDM systems where a single laser is used for many cores and modes \cite{6317137,6517220}. The sharing of lasers gives rise to laser phase noise which is highly correlated across the channels \cite{Lundberg2020,6517220,6317137}.

Various CPE methods that exploit such interchannel correlation have been investigated for multichannel systems. In master--slave CPE, estimates based on a single channel are used to compensate for the laser phase noise in all channels \cite{Lundberg2020,6517220,6317137}. This allows for complexity reduction in the digital signal processing (DSP) but does not improve the CPE performance. Another strategy is to perform joint-channel processing to obtain phase-noise estimates based on all channels, which improves the CPE performance at the potential cost of added DSP complexity. In \cite{6517220,Lundberg2020}, joint processing is implemented through phase averaging across the channels, which reduces the impact of additive noise that corrupts the phase-noise estimates. The drawback of phase-averaging and master--slave CPE is that these strategies implicitly rely on full phase-noise correlation across the channels to function effectively.

Laser phase noise is typically almost identical in the two polarizations of a four-dimensional (4D) channel, since both polarizations are generated by the same laser, but ensuring full phase-noise correlation across 4D channels can be challenging. This is due to the presence of other residual impairments such as nonlinear phase noise and frequency offsets \cite{8576586}, as well as optical interchannel delays that occur during propagation \cite{7328945,Lundberg2020}. In response to this, a pilot-aided algorithm is proposed for joint-channel CPE in \cite{8695027}. It can operate effectively for any interchannel correlation in the phase noise, but its performance is highly dependent on the pilot distribution over the channel--time symbol block.

The problem of identifying effective pilot distributions has been widely studied for channel estimation \cite{1312647,6585733,5419953,5946248}, as well as for joint CPE and channel estimation \cite{8108382}, in wireless orthogonal frequency-division multiplexing (OFDM) transmission.
In the context of fiber-optic communications, specific pilot distributions over the frequency--time grid in OFDM systems are proposed for I/Q-imbalance estimation in \cite{6236002}, as well as for the joint estimation of phase noise and other impairments in \cite{8052163,ZHANG201814}.
Focusing specifically on CPE performance, different pilot placement schemes are studied for single-channel transmission in \cite{4562696}. However, it is not clear from the literature which pilot distributions over the channel--time symbol block are effective for joint-channel CPE in multichannel fiber-optic transmission. This is particularly the case if arbitrary phase-noise correlation among the channels is considered.

In this paper, we study the problem of identifying effective pilot distributions over the channel--time symbol block. The pilots are used for joint-channel CPE in multichannel transmission impaired by laser phase noise.
We consider a simple multichannel phase-noise model, which makes no assumption on the phase-noise correlation across the channels and is used to develop the CPE algorithm in \cite[Sec.~III-B]{8695027}. The algorithm uses an extended Kalman smoother (EKS) to perform CPE jointly over all channels. The model and algorithm are experimentally verified in \cite{8576586}. In this paper, the model is particularized to the case where the phase noise is fully correlated over the two polarizations of each 4D channel, and arbitrarily but equally correlated across all 4D channels.
Using the considered model and algorithm, we formulate the problem of finding effective pilot distributions as a discrete optimization problem. We consider both the optimization of unstructured and structured pilot distributions. The optimization problems are solved using a genetic algorithm for different system parameters. We further consider several systematic constructions of heuristic pilot distributions that are compared with the optimized pilot distributions.

In \cite{alfredsson2017ecoc}, we presented preliminary results based on comparing several heuristic pilot distributions. This paper extends those results with the following contributions:
1) We show that there is negligible performance difference between unstructured and structured pilot distributions that have been numerically optimized. This implies that the parametrization of pilot distributions can be simplified without loss of performance;
2) We show that placing pilots identically (resulting in time-aligned pilots) in all channels or placing most pilots in a single channel are suboptimal strategies in general. Instead, distributing the pilots in a particular manner, referred to as \S{4} in the paper, attains the best results;
3) We extensively compare the use of time-aligned pilots versus \S{4} and show that for a fixed pilot rate, using \S{4} can substantially reduce the mean squared error (MSE) of the phase-noise estimates compared to using time-aligned pilots.
Furthermore, the achievable information rate (AIR) corresponding to each pilot distribution is maximized over the pilot rate.
We show that using \S{4} instead of time-aligned pilots can significantly increase the AIR for transmission of higher-order modulation formats.

\textit{Notation:}
Vectors and matrices are denoted by underlined letters $\underline x$ and uppercase sans-serif letters $\mathsf X$, respectively. The vector transpose is denoted by $(\cdot)^T$, and the trace of a square matrix is written as $\tr(\cdot)$. The Kronecker product is denoted by $\otimes$. Random quantities are denoted by boldface letters. The set of integers, real numbers, and complex numbers is denoted by $\mathbb Z$, $\mathbb R$, and $\mathbb C$, respectively. The imaginary unit is represented by $j$. The floor and ceiling functions are written as $\lfloor\cdot\rfloor$ and $\lceil\cdot\rceil$.

\section{System Model}
\label{sec:sys_model}

Consider uncoded dual-polarization transmission through $M/2$ 4D channels (frequency, space, or a combination thereof) through a coherent fiber-optic link, where each 4D channel comprises two complex channels. The total number of complex channels is therefore $M$. For brevity, complex channels will simply be referred to as channels throughout the rest of the paper. Blocks of $N$ complex symbols are transmitted in each channel, where each symbol either carries data or is a pilot. Data symbols are modelled as independent and identically distributed random variables that take on values in a zero-mean constellation $\mathcal X$ with equal probability. The constellation has average energy $E_\mathrm{s}$. Pilot symbols are modelled as random variables with a degenerate distribution, i.e., they have a probability 1 of being a complex point that is known to the transmitter and receiver. All pilots take on the same complex point $\zeta$, which is not necessarily in the constellation. The pilot distribution over the channel--time symbol block is known to both the transmitter and the receiver.

The received signal is assumed to have undergone a typical DSP chain \cite{7384692} that performs chromatic dispersion compensation, orthonormalization, timing recovery, adaptive equalization\footnote{The adaptive equalizer is assumed to be implemented in a phase-immune (blind or pilot-aided) fashion as in \cite{7384692}, which leaves the laser phase noise essentially unaffected.}, frame synchronization, frequency-offset compensation, and downsampling. The fiber Kerr nonlinearity is considered negligible. Assuming all DSP steps to have performed ideally, the processed signal is left with amplified spontaneous emission, modelled as additive white Gaussian noise (AWGN), and laser phase noise. Considering one sample per symbol, the received and processed signal in the $i$th channel is thus described in complex baseband at time $k$ as
\begin{equation}
	\bm r_{i,k}=\bm s_{i,k}e^{j\bm\theta_{i,k}}+\bm n_{i,k},
\end{equation}
for $k=1,\dots,N$ and $i=1,\dots,M$, where $\bm r_{i,k}$, $\bm s_{i,k}$, and $\bm n_{i,k}$ are the received and processed samples, transmitted symbols, and AWGN samples, respectively. The AWGN is assumed to have the same variance in all channels, i.e., $N_0/2$ per real dimension. Moreover, $\underline{\bm r}_k=[\bm r_{1,k},\dots,\bm r_{M,k}]^T$, with $\underline{\bm s}_k$ and $\underline{\bm n}_k$ being defined similarly. The channel--time symbol block over which pilots are distributed is thus encapsulated in an $M\times N$ matrix.

The laser phase noise $\underline{\bm\theta}_k=[\bm\theta_{1,k},\dots,\bm\theta_{M,k}]^T$ is modelled jointly over all channels as a multidimensional Gaussian random walk, described as
\begin{equation}
	\underline{\bm\theta}_k=\underline{\bm\theta}_{k-1}+\dot{\underline{\bm\theta}}_k,
	\label{eq:pn_rw}
\end{equation}
where $\dot{\underline{\bm\theta}}_k$ is a multivariate zero-mean Gaussian random variable with covariance matrix $\mathsf Q$ and $\underline{\bm\theta}_1$ is uniformly distributed on $[0,2\pi)^M$. Furthermore, $\mathsf Q$ describes the phase-noise correlation across the channels. Full phase-noise correlation over the two polarizations in all 4D channels is considered. Moreover, as already mentioned, the phase noise can be arbitrarily correlated across 4D channels depending on the system. For simplicity, the same correlation across all 4D channels is assumed. Hence, $\mathsf Q$ is parameterized in terms of the combined laser linewidth of the system, $\Delta\nu$, and the 4D-channel correlation, $\alpha\in[0,1]$, where $\alpha=0$ gives uncorrelated phase noise and $\alpha=1$ gives fully correlated phase noise across the 4D channels.
The matrix $\mathsf Q$ is expressed as
\begin{equation}
	\mathsf Q=\begin{bmatrix}
		\mathsf J_2 & \alpha\mathsf J_2 & \alpha\mathsf J_2 & \cdots & \alpha\mathsf J_2 \\
		\raisebox{2pt}{$\alpha\mathsf J_2$} & \raisebox{2pt}{$\mathsf J_2$} & & & \vdots \\[-2pt]
		\raisebox{2pt}{$\alpha\mathsf J_2$} & & \ddots & & \vdots \\[-2pt]
		\vdots & & & \ddots & \raisebox{2pt}{$\alpha\mathsf J_2$} \\[2.7pt]
		\alpha\mathsf J_2 & \cdots & \cdots & \alpha\mathsf J_2 & \mathsf J_2
	\end{bmatrix},\label{eq:rwcov}
\end{equation}
where $\mathsf J_2$ is a $2\times2$ matrix of ones.

\section{Pilot Distributions}
\label{sec:pdist_opt}

In this section, we formulate the problem of determining effective pilot distributions as a discrete optimization problem. In addition, we proposed several systematic pilot-distribution constructions.

The pilot-distribution optimization over the channel--time symbol block is carried out by minimizing the MSE of the phase-noise estimates, averaged over the channels. An algorithm proposed in \cite[Sec.~III-B]{8695027} is utilized, which performs iterative joint-channel CPE and data detection. It was shown in \cite{8695027} that the algorithm outperforms the blind phase search algorithm \cite{4814758} for transmission through a single channel, and that its performance improves with the number of channels. The joint-channel CPE is carried out using an EKS, which entails forward--backward recursions to produce estimates based on all available received samples, i.e., $\underline{\bm r}_1,\dots,\underline{\bm r}_{N}$. In the forward recursion, filtered phase-noise estimates are obtained, which is followed by a backward recursion that yields smoothed phase-noise estimates.

A typical recursive-filtering notation will be adopted in what follows. The subindex $k|k-1$ is used to denote matrices corresponding to estimates at time $k$ based on $\underline{\bm r}_1,\dots,\underline{\bm r}_{k-1}$, where $k$ is the time index. Similarly, $k|k$ is used for matrices corresponding to estimates at time $k$ based on $\underline{\bm r}_1,\dots,\underline{\bm r}_{k}$, and $k|N$ is used for matrices corresponding to estimates at time $k$ based on all samples, $\underline{\bm r}_1,\dots,\underline{\bm r}_{N}$.
The MSE of the resulting smoothed estimates is encapsulated in the diagonal elements of the covariance matrix $\mathsf{M}_{k|N}$, which is computed for all $k$ by the EKS through the recursive equations
\begin{align}
	\mathsf{M}_{k|k-1}&=\mathsf{M}_{k-1|k-1}+\mathsf Q,\\
	\mathsf{M}_{k|k}&=\left(\mathsf{I}+\mathsf{M}_{k|k-1}\mathsf{V}_k\right)^{-1}\mathsf{M}_{k|k-1},
\end{align}
for $k=2,3,\dots,N$, followed by
\begin{align}
	\mathsf{A}_k&=\mathsf{M}_{k|k}\left(\mathsf{M}_{k+1|k}\right)^{-1},\\
	\mathsf{M}_{k|N}&=\mathsf{M}_{k|k}+\mathsf{A}_k\left(\mathsf{M}_{k+1|N}-\mathsf{M}_{k+1|k}\right)\mathsf{A}_k^T,
	\label{eq:Ms}
\end{align}
for $k=N-1,N-2,\dots,1$. Moreover,
\begin{equation}
	\mathsf{V}_k=\diag\left(\frac{|\tilde s_{1,k}|^2}{\tilde\sigma_{1,k}^2},\dots,\frac{|\tilde s_{M,k}|^2}{\tilde\sigma_{M,k}^2}\right),
\end{equation}
where $\diag(\cdot)$ denotes a diagonal matrix, $\tilde s_{i,k}=\zeta$ and $\tilde\sigma_{i,k}^2=N_0/2$ for pilots, whereas for data symbols, $\tilde s_{i,k}=0$ and $\tilde\sigma_{i,k}^2=(N_0+E_\mathrm{s})/2$.
The forward recursions are initialized with $\mathsf{M}_{1|1}=\diag(\tilde\sigma_{1,1}^2/E_\mathsf{s},\dots,\tilde\sigma_{M,1}^2/E_\mathsf{s})$.

In the first iteration of the algorithm in \cite{8695027}, the EKS estimates the phase noise across the channel--time symbol block by interpolating the estimated phase noise between the pilots, jointly over all channels. We point interested readers to \cite{8695027} for the derivation of the algorithm, but warn about notational differences between this paper and \cite{8695027}.

\subsection{Unstructured Optimization}
\label{sec:pdist_opt_unstr}

Without imposing any constraints on its structure, a pilot distribution denoted by \U{opt} can be parameterized by $\underline p_\mathrm{u}=[p_1,\dots,p_L]$, where $p_l$ describes the position of the $l$th pilot in the channel--time symbol block. As there are $MN$ slots in such a block, these positions range from $1$ to $MN$. Each position $p_l$ is then mapped to the index $(i,k)$ in the channel--time symbol block as $i=\mathrm{mod}(p_l-1,M)+1$ and $k=\lceil p_l/M\rceil$. Moreover, $L/(MN)$ gives the pilot rate, averaged over the channels. A discrete optimization problem is thus formulated as
\begin{equation}
	\min_{\underline p_\mathrm{u}\in\mathbb Z^L}\sum_{k=1}^{N}\tr\left(\mathsf{M}_{k|N}\right),
	\label{eq:opt_u}
\end{equation}
subject to $1\leq p_l\leq MN$ for all $l=1,\dots,L$. The quantity that is minimized in \eqref{eq:opt_u} is proportional to the MSE of the phase-noise estimates, averaged over the channels.

\subsection{Structured Optimization}
\label{sec:pdist_opt_str}

The optimization in \eqref{eq:opt_u} quickly becomes infeasible to carry out for large $M$ and $N$. To alleviate this, an additional structure is introduced to the optimization problem. Consider a pilot distribution denoted by \S{opt}, which entails a pilot in the initial symbol slot of each channel. This is done since the random walk in \eqref{eq:pn_rw} is initialized with $\bm\theta_1$ uniformly distributed on $[0,2\pi)^M$. The initial pilot in the $i$th channel is followed by a sequence of $\kappa-1$ equispaced pilots with spacing $\tau_i$, starting at the $\delta_i$th position in the symbol block. The resulting pilot rate is $\kappa/N$. The distribution is thus parameterized by $2M$ variables $\underline p_\mathrm{s}=[\delta_1,\tau_1,\delta_2,\dots,\delta_M,\tau_M]$ and an optimization problem is formulated as
\begin{equation}
	\min_{\underline p_\mathrm{s}\in\mathbb Z^{2M}}\sum_{k=1}^N\tr\left(\mathsf{M}_{k|N}\right),
	\label{eq:opt_s}
\end{equation}
subject to $\delta_i\geq2$, $\tau_i\geq1$, and $\delta_i+\tau_i(\kappa-1)\leq N$ for all $i=1,\dots,M$. The optimization problems in \eqref{eq:opt_u} and \eqref{eq:opt_s} are carried out using a genetic algorithm implemented in Matlab's global-optimization toolbox.

\subsection{Heuristic Pilot Distributions}
\label{sec:heur_pdists}

In addition to the optimized pilot distributions, several systematic constructions of heuristic pilot distributions are considered and denoted by \S{1}--\S{5}. All the constructions depend solely on $\kappa$, $M$, and $N$, and the resulting pilot rate is $\kappa/N$ when averaged over all channels. These distributions are illustrated in Fig.~\ref{fig:heur_pdists} and their constructions are detailed in Appendix \ref{app:heur_constr}. Note that similarly to \S{opt}, a pilot is placed at the initial symbol slot of each channel due to the same reason as the one mentioned in Sec.~\ref{sec:pdist_opt_str}.

\begin{figure*}[!t]
	\centering
	\includegraphics{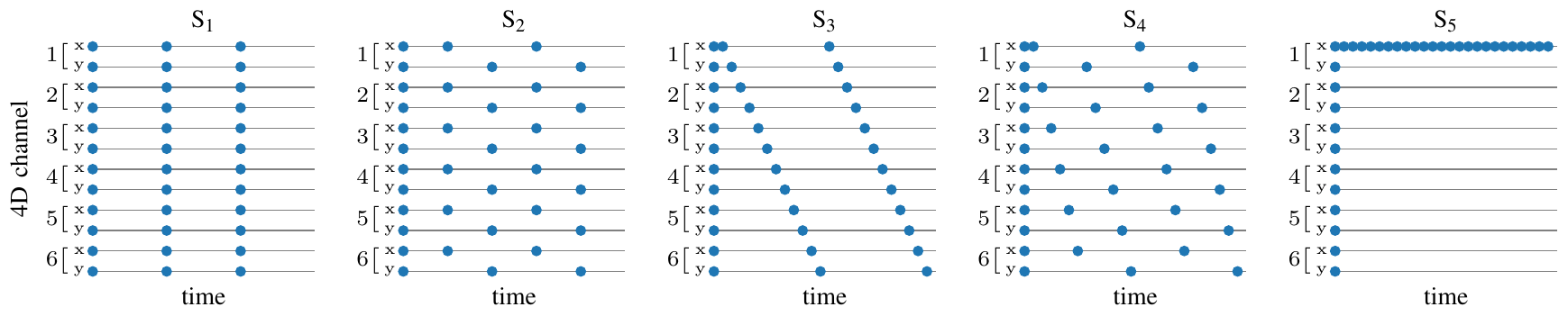}
	\caption{Illustration of the considered structured pilot distributions for transmission through six 4D channels, where each 4D channel comprises $\mathrm x$ and $\mathrm y$ polarizations.}
	\label{fig:heur_pdists}
\end{figure*}

An unstructured distribution that will also be considered for reference is denoted by \U{rnd} and entails randomized pilot placements, in which a total of $\kappa$ pilots are placed per channel using random sampling without replacement \cite[Ch.~2]{johnson:sampling:2012}.

\begin{figure}[!t]
	\centering
	\includegraphics{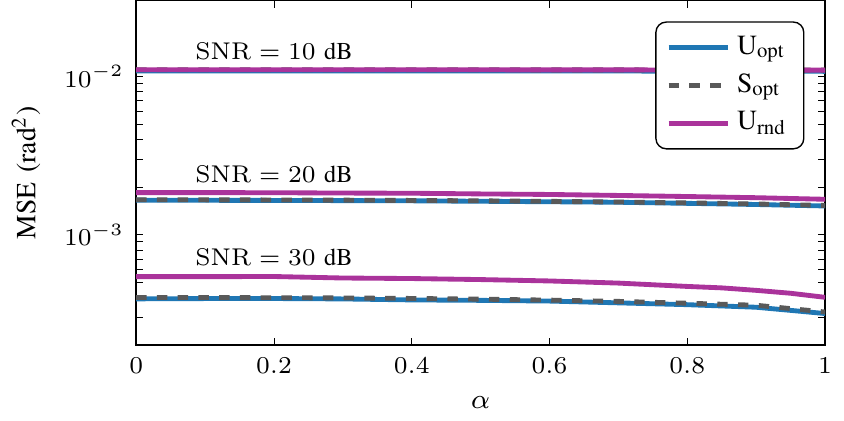}
	\caption{MSE vs. $\alpha$ for \U{opt}, \S{opt}, and \U{rnd} for 5\% pilot rate (100 symbols per channel, 20 pilots in total across all channels).}
	\label{fig:MSE_opt_unc_vs_str}
\end{figure}
 
\section{Numerical Results}
\label{sec:results}

In this section, the pilot distributions are assessed based on the MSE of the phase-noise estimates as well as on the AIR of the system. Unless otherwise specified, blocks of $N=10\,000$ symbols per channel are transmitted. Pilots take on the point $\zeta=\sqrt{E_\mathrm{s}}$. To compute the MSE, a single transmission suffices as $\mathsf{M}_{k|N}$ in \eqref{eq:Ms} is deterministically computed given a pilot distribution. Each channel is modulated independently, and the considered complex modulation formats are 64-ary quadrature amplitude modulation (64QAM), 256QAM, and 1024QAM with Gray-labeled symbols. The AIR is quantified in terms of bits per complex symbol. It is obtained by estimating the generalized mutual information\footnote{The LLRs are obtained by the algorithm in \cite[Sec.~III-B]{8695027}.} (GMI) \cite[Eq.~(36)]{Alvarado:18} and accounting for rate loss due to the pilot insertions. The GMI is estimated using Monte Carlo simulations, where random symbol blocks are repeatedly generated and transmitted until statistically reliable AIR estimates are obtained. SNRs in the range 10 dB to 40 dB are considered, where for reference an SNR of 19.73 dB, 25.43 dB, and 31.11 dB gives a theoretical bit error probability of 0.01 for 64QAM, 256QAM, and 1024QAM, respectively, for uncoded transmission over the AWGN channel \cite[Eq.~(17)]{1021039}. Finally, unless otherwise specified, a total linewidth of $\Delta\nu=200$ kHz and 20 GBd symbol rate are considered, which are parameter values commonly seen in experimental demonstrations \cite{Lundberg2020,Mazur:19,8576586}.

\subsection{MSE Results}
\label{sec:mse_results}

Fig.~\ref{fig:MSE_opt_unc_vs_str} compares the MSEs corresponding to \U{opt} and \S{opt} vs. 4D-channel correlation $\alpha$ for $M=4$ and different SNRs, with 100 symbols in each channel. The optimization is carried out for each tested set of parameter values. A total of 20 pilots is used, resulting in 5\% pilot rate. The average ensemble performance of \U{rnd} based on 1000 realizations is also included for reference. In general, the two optimization strategies yield similar pilot distributions, and hence similar MSE results for all tested SNRs and values of $\alpha$. More specifically, \S{opt} gives approximately 1.4\% higher MSE than \U{opt} on average. This indicates that the optimization of pilot distributions can be simplified by introducing structure without a significant loss in optimality. Consequently, all optimization results in the remainder of this paper will correspond to \S{opt}.

\begin{figure}[!t]
	\centering
	\includegraphics{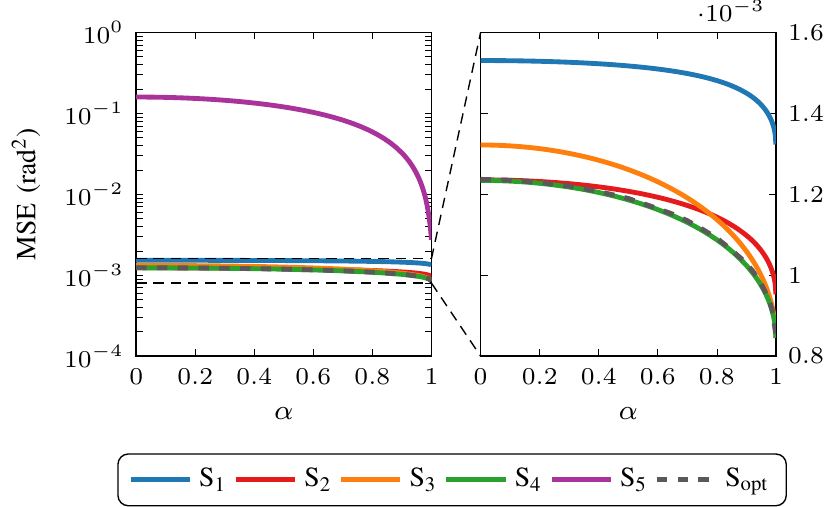}
	\caption{MSE vs. 4D-channel correlation $\alpha$ for the considered heuristic pilot distributions with 1\% pilot rate. The right plot shows a zoomed-in version of the same results.}
	\label{fig:MSE_ali_1ch_vs_corr}
\end{figure}

Fig.~\ref{fig:MSE_ali_1ch_vs_corr}
shows MSE vs. 4D-channel correlation $\alpha$ comparing \S{1}--\S{5} and \S{opt} for $M=4$, 1\% pilot rate, and an SNR of 25 dB. Placing pilots according to \S{5}, which entails inserting essentially all the available pilots in a single channel, is highly suboptimal, particularly at low values of $\alpha$. This is because it relies on fully correlated channels to work properly, similarly to master--slave estimation or self-homodyne detection \cite{Lundberg2020,7328945}. Moreover, \S{5} does not attain the performance of \S{opt} at $\alpha=1$. This is due to the AWGN corrupting the estimates based on the pilots in the first column of the channel--time symbol block. Placing pilots in more columns can improve the performance of \S{5}, however, provided that $\alpha=1$.

\begin{figure}[!t]
	\centering
	\includegraphics{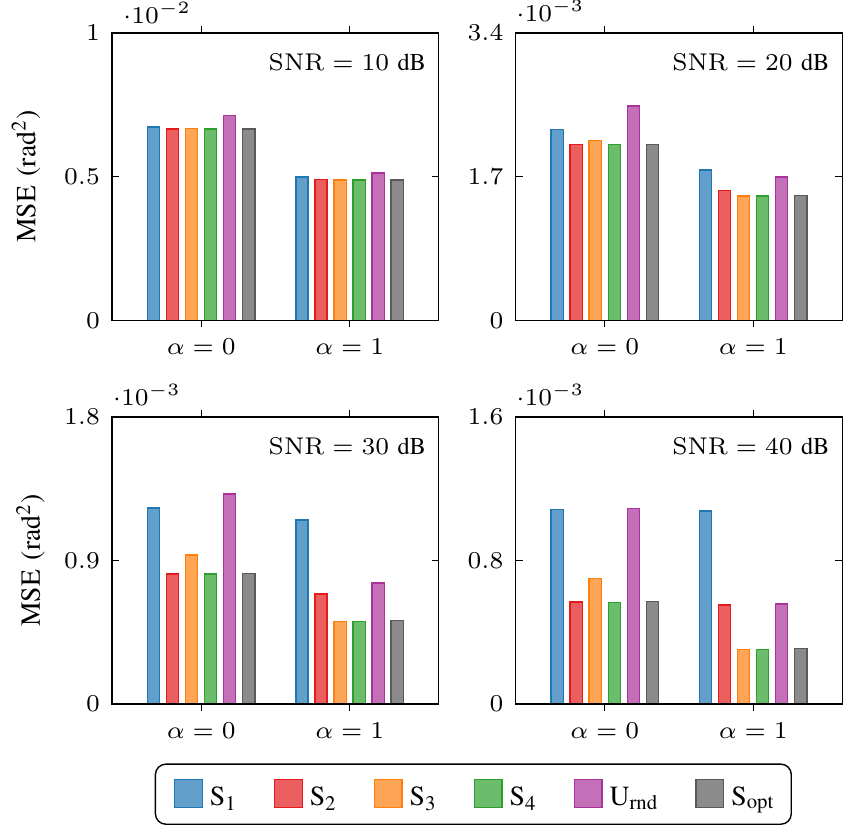}
	\caption{MSE results comparing several pilot distributions for 1\% pilot rate but different values of SNR and 4D-channel correlation $\alpha$.}
	\label{fig:MSE_opt_vs_heur}
\end{figure}

Placing pilots according to \S{1}, i.e., using identical placements in all channels resulting in time-aligned pilots, is suboptimal in general. This is because the two polarizations in each 4D channel are fully correlated, and hence have identical phase noise. Therefore, it is more effective to position the pilots differently in the two polarizations, such as in \S{2}. Indeed, \S{2} attains the optimized MSE performance at $\alpha=0$ but becomes suboptimal as $\alpha$ increases, as it does not exploit the correlation among 4D channels effectively.

\S{3} attains the performance of \S{opt} for $\alpha=1$ since the pilots are placed in a cyclic-shift pattern across the 4D channels. However, it becomes suboptimal as $\alpha$ decreases. The reason for this becomes clear if $\alpha=0$ is considered, in which case the algorithm resorts to independent joint-polarization processing in each 4D channel. When 4D channels are treated independently, \S{3} is similar to \S{1} in the sense that the pilots are not spread effectively over the two polarizations. \S{4} attains the performance of \S{opt} for all values of $\alpha$. This is by virtue of the effective pilot spreading over the two polarizations of each 4D channel, as in \S{2}, as well as the cyclic-shift pilot pattern that is implemented across the 4D channels, as in \S{3}.

\begin{figure}[!t]
	\centering
	\includegraphics{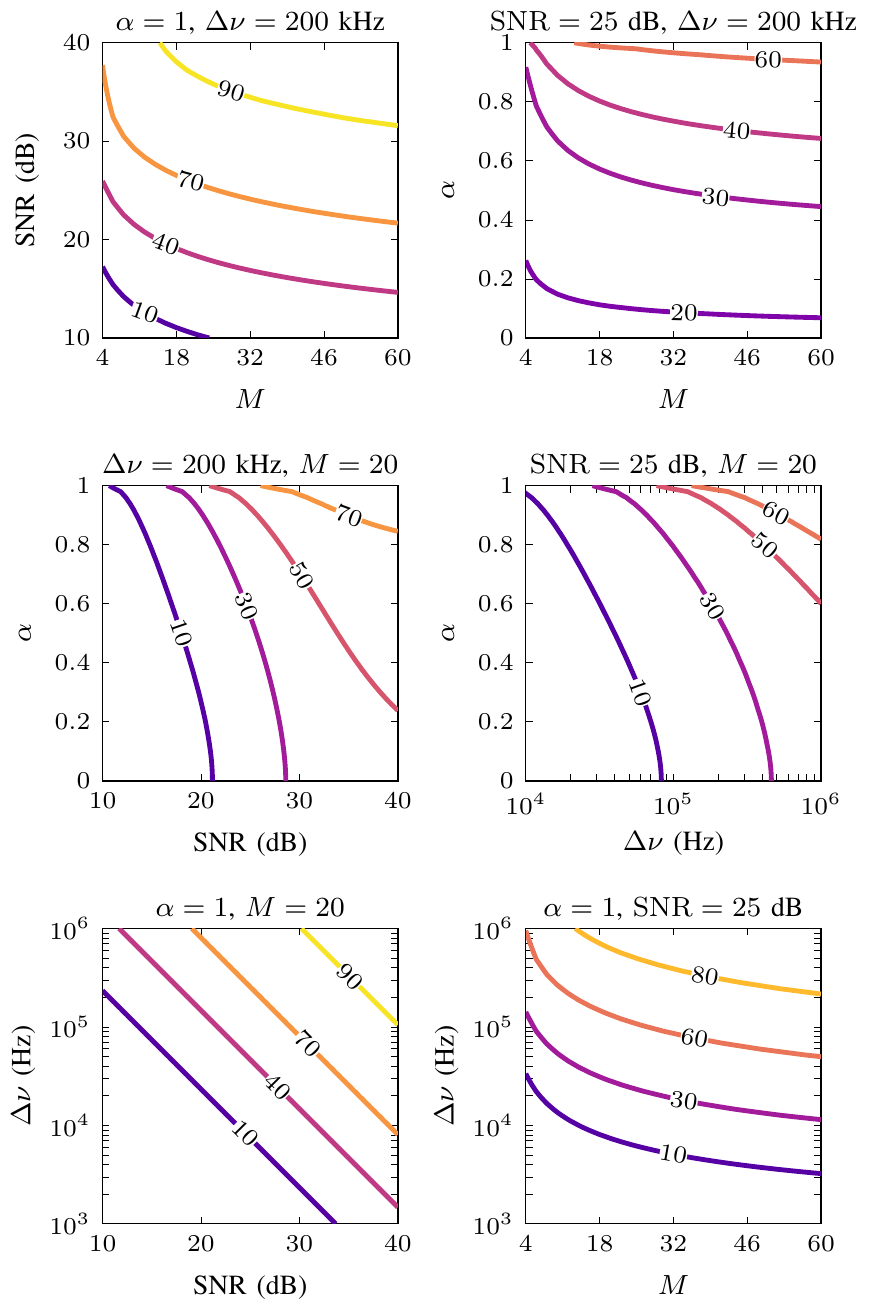}
	\caption{The reduction in MSE that results by placing pilots according to \S{4} instead of \S{1} (time-aligned pilots).}
	\label{fig:MSEreduct_vs_params}
\end{figure}

\begin{table*}[!t]
	\renewcommand{\arraystretch}{1.5}
	\caption{AIR gain in bits per complex symbol by placing pilots according to \S{4} as opposed to using time-aligned pilots (\S{1}) for $\Delta\nu=200$ kHz. The AIRs for both pilot distributions are maximized over the pilot rate before comparison is made.}
	\label{table:AIR_gain}
	\centering
	\begin{tabular}{c|c|ccc|c|ccc|c|ccc}
		\hline\hline
		& \multicolumn{4}{c|}{AIR gain for 64QAM (b/sym)} & \multicolumn{4}{c|}{AIR gain for 256QAM (b/sym)} & \multicolumn{4}{c}{AIR gain for 1024QAM (b/sym)} \\ \cline{2-13}
		& SNR & $\alpha=0$ & $\alpha=0.5$ & $\alpha=1$ & SNR & $\alpha=0$ & $\alpha=0.5$ & $\alpha=1$ & SNR & $\alpha=0$ & $\alpha=0.5$ & $\alpha=1$ \\ \hline
		\multirow{3}{*}{$M=4$} & $15$ dB & $0$ & $0$ & $0.01$ & $20$ dB & $0.01$ & $0.01$ & $0.02$ & $25$ dB & $0.03$ & $0.03$ & $0.07$ \\
		& $20$ dB & $0.01$ & $0.01$ & $0.02$ & $25$ dB & $0.03$ & $0.03$ & $0.06$ & $30$ dB & $0.09$ & $0.09$ & $0.16$ \\
		& $25$ dB & $0.02$ & $0.02$ & $0.02$ & $30$ dB & $0.05$ & $0.05$ & $0.08$ & $35$ dB & $0.13$ & $0.15$ & $0.24$ \\ \hline
		\multirow{3}{*}{$M=40$} & $15$ dB & $0$ & $0.01$ & $0.03$ & $20$ dB & $0.01$ & $0.02$ & $0.07$ & $25$ dB & $0.03$ & $0.06$ & $0.18$ \\
		& $20$ dB & $0.01$ & $0.02$ & $0.05$ & $25$ dB & $0.03$ & $0.05$ & $0.16$ & $30$ dB & $0.08$ & $0.14$ & $0.36$ \\
		& $25$ dB & $0.01$ & $0.02$ & $0.04$ & $30$ dB & $0.05$ & $0.07$ & $0.14$ & $35$ dB & $0.13$ & $0.19$ & $0.41$\\ \hline\hline
	\end{tabular}
\end{table*}

Fig.~\ref{fig:MSE_opt_vs_heur} gives a further comparison between \S{1}--\S{4} and \S{opt}, as well as the average ensemble performance of \U{rnd} based on 1000 realizations, in terms of MSE for $M=4$, 1\% pilot rate, $\alpha=\{0,1\}$, and several SNRs.
The results show that at lower SNRs, the choice of pilot distribution is less important than at higher SNRs. More specifically, at an SNR of 10 dB, there is a negligible MSE difference between the considered distributions, regardless of $\alpha$. At higher SNRs, however, the difference becomes substantial. Furthermore, a randomized pilot distribution with no particular structure has similar or better performance on average than \S{1} in most of the tested cases. However, it does not attain the performance of \S{opt} in any case.

\S{1} is arguably the most typical choice out of the considered heuristic distributions, but as is shown in Figs.~\ref{fig:MSE_ali_1ch_vs_corr} and \ref{fig:MSE_opt_vs_heur}, it is suboptimal in general for joint-channel CPE. In contrast, \S{4} attains the performance of \S{opt} for all tested parameters. Fig.~\ref{fig:MSEreduct_vs_params} presents the MSE reduction that results from using \S{4} instead of \S{1} for different parameter values and 1\% pilot rate. For low values of $\Delta\nu$ and SNR, using \S{4} gives less than 10\% MSE reduction, implying that the choice of pilot distribution gives a marginal performance difference. However, this choice becomes more impactful with increasing values of $\alpha$, SNR, $\Delta\nu$, and $M$, yielding more than 90\% reduction in MSE in some cases. 

\begin{figure}[!t]
	\centering
	\includegraphics{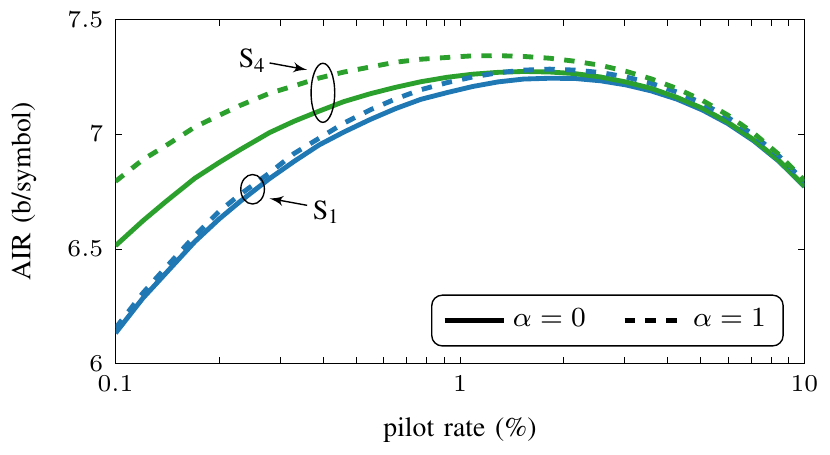}
	\caption{AIR vs. pilot rate for transmission of 256QAM over two 4D channels at SNR of 25 dB as a function of pilot rate for different pilot distributions.}
	\label{fig:AIR_vs_prate}
\end{figure}

\subsection{AIR Results}
\label{sec:air_results}

It is clear from the results shown so far that the CPE can be improved through effective pilot distributions over the channel--time symbol block. Hence, the CPE performance can be improved for a fixed pilot rate, or alternatively, the pilot rate can be reduced while maintaining the same CPE performance. This can be exploited to increase the AIR of the system.

Fig.~\ref{fig:AIR_vs_prate} illustrates this concept, where the AIR is estimated as a function of the pilot rate for transmission of 256QAM at an SNR of 25 dB, comparing \S{1} and \S{4} for two values of $\alpha$. The AIR curves have a peak corresponding to an optimal pilot rate, similarly to what was found in \cite{Mazur:19}. This is because too low pilot rates result in a crude CPE, which leads to a reduction in the AIR. Increasing the number of pilots improves the CPE performance up to a certain point, but eventually the corresponding rate reduction due to the pilot insertion begins to outweigh the CPE improvements.

Maximizing the AIRs corresponding to \S{1} and \S{4} over the pilot rate, while keeping other system parameters fixed, gives an estimated AIR gain that results from using \S{4} instead of \S{1}. When \S{1} (respectively, \S{4}) is used, the optimal pilot rates are found to range from 0.5\% to 3\% (respectively, 0.1\% to 2.3\%) for the tested system parameters. Table \ref{table:AIR_gain} summarizes the resulting AIR gain for different modulation formats, SNRs, and values of $\alpha$ and $M$. The gain increases in general with the modulation order, $M$, and $\alpha$. In particular, for 64QAM, a marginal gain of up to 0.05 b/sym is found. However, for 256QAM and 1024QAM respectively, an AIR increase of up to 1.1\% (from 7.80 b/sym to 7.88 b/sym) and 2.5\% (from 9.44 b/sym to 9.67 b/sym) is observed for $M=4$, as well as an increase of up to 2.2\% (from 7.32 b/sym to 7.48 b/sym) and 4.4\% (from 9.45 b/sym to 9.86 b/sym) for $M=40$. Hence, in cases where phase noise is the limiting performance factor, the choice of pilot distribution has a significant impact on the information rate of the system.

\section{Conclusions}
\label{sec:concl}

In multichannel transmission where the laser phase noise is correlated across channels, pilot-aided joint-channel carrier-phase estimation based on extended Kalman smoothing has proven to be effective. However, the choice of pilot distribution over the time--channel symbol block can have a strong impact on the resulting performance.
In this paper, the problem of identifying effective pilot distributions was formulated as a discrete optimization problem. The considered multichannel model entailed laser phase noise that is fully correlated over the two polarizations in each 4D channel, and arbitrarily but equally correlated across the 4D channels. Using this model, optimized pilot distributions were found via minimizing the mean squared error of the phase-noise estimates that were obtained using extended Kalman smoothing. In addition, several heuristic pilot distributions were proposed. These distributions were extensively compared for different system parameters.

Based on the optimization results, it was shown that it is suboptimal to place pilots identically in all channels or to place most of the pilots in a single channel (see Figs.~\ref{fig:MSE_opt_unc_vs_str} and \ref{fig:MSE_ali_1ch_vs_corr}). Instead, placing pilots on a particular grid (see \S{4} in Fig.~\ref{fig:heur_pdists}) was found to attain the best performance. In particular, the choice of pilot distribution becomes more impactful with increasing SNR, 4D-channel correlation, number of channels, and laser linewidth (see Figs.~\ref{fig:MSE_opt_vs_heur} and \ref{fig:MSEreduct_vs_params}). Finally, the AIR for transmission of higher-order QAM formats was maximized over the pilot rate for the considered pilot distributions. It was observed that the choice of pilot distribution can considerably affect the AIR for transmission of higher-order modulation formats (see Table \ref{table:AIR_gain}).

Accounting for the effects of nonlinear phase noise, using a different pilot-aided CPE algorithm, or jointly optimizing pilot distributions for pilot-aided adaptive equalization and carrier recovery may lead to different optimization results. Such extensions to the considered problem are left for future work.

\appendices
\section{Systematic Pilot-Distribution Constructions}
\label{app:heur_constr}
This section explains the construction of the pilot distributions in  Fig.~\ref{fig:heur_pdists}. \S{1}--\S{4} are parameterized analogously to \S{opt} in Section \ref{sec:pdist_opt_str} using $\delta_i$ and $\tau_i$ for $i=1,\dots,M$.

\S{1} is defined for $0\leq\kappa\leq N$. It is constructed using $\tau_i=N/\kappa$ and $\delta_i=1+\tau_i$ for all $i$.
\S{2} is defined for $0\leq\kappa\leq N/2$. It is constructed using $\tau_i=N/(\kappa-1/2)$ for all $i$, $\delta_i=1+\tau_i$ if $i$ is even, and $\delta_i=1+\tau_i/2$ if $i$ is odd.
\S{3} is defined for $0\leq\kappa\leq N/M$. It is constructed using $\tau_i=N/(\kappa-1+1/M)$ and $\delta_i=1+i\tau_i/M$ for all $i$.
\S{4} is defined for $0\leq\kappa\leq N/M$. It is constructed using $\tau_i=N/(\kappa-1+1/M)$ and $\delta_i=1+v_i\tau_i/M$ for all $i$, where $v_i=(2i+(M-1)(-1)^i+M+1)/4$.
\S{5} is defined for $0\leq\kappa\leq N/M$. It consists of $M\kappa-M+1$ pilots in one channel with a spacing $N/(M\kappa-M+1)$ and initial position $1$, as well as one pilot in the initial symbol slot of the other $M-1$ channels.

In the case that the pilot positions computed from the above constructions are not integers, they are rounded to the nearest integer.



\begin{thebibliography}{10}
	\providecommand{\url}[1]{#1}
	\csname url@samestyle\endcsname
	\providecommand{\newblock}{\relax}
	\providecommand{\bibinfo}[2]{#2}
	\providecommand{\BIBentrySTDinterwordspacing}{\spaceskip=0pt\relax}
	\providecommand{\BIBentryALTinterwordstretchfactor}{4}
	\providecommand{\BIBentryALTinterwordspacing}{\spaceskip=\fontdimen2\font plus
		\BIBentryALTinterwordstretchfactor\fontdimen3\font minus
		\fontdimen4\font\relax}
	\providecommand{\BIBforeignlanguage}[2]{{%
			\expandafter\ifx\csname l@#1\endcsname\relax
			\typeout{** WARNING: IEEEtran.bst: No hyphenation pattern has been}%
			\typeout{** loaded for the language `#1'. Using the pattern for}%
			\typeout{** the default language instead.}%
			\else
			\language=\csname l@#1\endcsname
			\fi
			#2}}
	\providecommand{\BIBdecl}{\relax}
	\BIBdecl
	
	\bibitem{8346089}
	R.~Maher, K.~Croussore, M.~Lauermann, R.~Going, X.~Xu, and J.~Rahn,
	``Constellation shaped 66 {GB}d {DP-1024QAM} transceiver with 400 km
	transmission over standard {SMF},'' in \emph{Proc. European Conf. Opt.
		Commun.}, Sep. 2017, p. Th.PDP.B.2.
	
	\bibitem{6886985}
	S.~{Beppu}, K.~{Kasai}, M.~{Yoshida}, and M.~{Nakazawa}, ``{2048 QAM (66
		Gbit/s)} single-carrier coherent optical transmission over 150 km with a
	potential {SE} of 15.3 bit/s/{Hz},'' in \emph{Proc. Opt. Fiber Commun.
		Conf.}, Mar. 2014.
	
	\bibitem{Olsson:18}
	S.~L. Olsson, J.~Cho, S.~Chandrasekhar, X.~Chen, P.~J. Winzer, and S.~Makovejs,
	``Probabilistically shaped {PDM 4096-QAM} transmission over up to 200 km of
	fiber using standard intradyne detection,'' \emph{Opt. Express}, vol.~26,
	no.~4, pp. 4522--4530, Feb. 2018.
	
	\bibitem{2019:chen}
	X.~Chen, J.~Cho, A.~Adamiecki, and P.~Winzer, ``16384-{QAM} transmission at 10
	{GBd} over 25-km {SSMF} using polarization-multiplexed probabilistic
	constellation shaping,'' in \emph{Proc. European Conf. Opt. Commun.}, Sep.
	2019, p. PD.3.3.
	
	\bibitem{5447711}
	M.~{Kuschnerov}, M.~{Chouayakh}, K.~{Piyawanno}, B.~{Spinnler}, E.~{de Man},
	P.~{Kainzmaier}, M.~S. {Alfiad}, A.~{Napoli}, and B.~{Lankl}, ``Data-aided
	versus blind single-carrier coherent receivers,'' \emph{{IEEE} Photon. J.},
	vol.~2, no.~3, pp. 387--403, Jun. 2010.
	
	\bibitem{4814758}
	T.~Pfau, S.~Hoffmann, and R.~No{\'e}, ``Hardware-efficient coherent digital
	receiver concept with feedforward carrier recovery for {$M$-QAM}
	constellations,'' \emph{J. Lightw. Technol.}, vol.~27, no.~8, pp. 989--999,
	Apr. 2009.
	
	\bibitem{4298982}
	E.~Ip and J.~M. Kahn, ``Feedforward carrier recovery for coherent optical
	communications,'' \emph{J. Lightw. Technol.}, vol.~25, no.~9, pp. 2675--2692,
	Sep. 2007.
	
	\bibitem{Mazur:19}
	M.~Mazur, J.~Schr\"{o}der, A.~Lorences-Riesgo, T.~Yoshida, M.~Karlsson, and
	P.~A. Andrekson, ``Overhead-optimization of pilot-based digital signal
	processing for flexible high spectral efficiency transmission,'' \emph{Opt.
		Express}, vol.~27, no.~17, pp. 24\,654--24\,669, Aug. 2019.
	
	\bibitem{8859308}
	T.~Sasai, A.~Matsushita, M.~Nakamura, S.~Okamoto, F.~Hamaoka, and Y.~Kisaka,
	``Laser phase noise tolerance of uniform and probabilistically shaped {QAM}
	signals for high spectral efficiency systems,'' \emph{J. Lightw. Technol.},
	vol.~38, no.~2, pp. 439--446, Jan. 2020.
	
	\bibitem{8695027}
	A.~F. {Alfredsson}, E.~{Agrell}, and H.~{Wymeersch}, ``Iterative detection and
	phase-noise compensation for coded multichannel optical transmission,''
	\emph{{IEEE} Trans. Commun.}, vol.~67, no.~8, pp. 5532--5543, Aug. 2019.
	
	\bibitem{7384692}
	D.~S. Millar, R.~Maher, D.~Lavery, T.~Koike-Akino, M.~Pajovic, A.~Alvarado,
	M.~Paskov, K.~Kojima, K.~Parsons, B.~C. Thomsen, S.~J. Savory, and P.~Bayvel,
	``Design of a 1 {Tb/s} superchannel coherent receiver,'' \emph{J. Lightw.
		Technol.}, vol.~34, no.~6, pp. 1453--1463, Mar. 2016.
	
	\bibitem{7301998}
	M.~P. Yankov, T.~Fehenberger, L.~Barletta, and N.~Hanik, ``Low-complexity
	tracking of laser and nonlinear phase noise in {WDM} optical fiber systems,''
	\emph{J. Lightw. Technol.}, vol.~33, no.~23, pp. 4975--4984, Dec. 2015.
	
	\bibitem{8327487}
	M.~Mazur, A.~Lorences-Riesgo, J.~Schr\"{o}der, P.~A. Andrekson, and
	M.~Karlsson, ``10 {Tb/s PM-64QAM} self-homodyne comb-based superchannel
	transmission with 4\% shared pilot tone overhead,'' \emph{J. Lightw.
		Technol.}, vol.~36, no.~16, pp. 3176--3184, Aug. 2018.
	
	\bibitem{6317137}
	M.~D. Feuer, L.~E. Nelson, X.~Zhou, S.~L. Woodward, R.~Isaac, B.~Zhu, T.~F.
	Taunay, M.~Fishteyn, J.~M. Fini, and M.~F. Yan, ``Joint digital signal
	processing receivers for spatial superchannels,'' \emph{{IEEE} Photon.
		Technol. Lett.}, vol.~24, no.~21, pp. 1957--1960, Nov. 2012.
	
	\bibitem{6517220}
	R.~G.~H. van Uden, C.~M. Okonkwo, V.~A. J.~M. Sleiffer, M.~Kuschnerov,
	H.~de~Waardt, and A.~M.~J. Koonen, ``Single {DPLL} joint carrier phase
	compensation for few-mode fiber transmission,'' \emph{{IEEE} Photon. Technol.
		Lett.}, vol.~25, no.~14, pp. 1381--1384, Jul. 2013.
	
	\bibitem{Lundberg2020}
	L.~Lundberg, M.~Mazur, A.~Mirani, B.~Foo, J.~Schr{\"o}der, V.~Torres-Company,
	M.~Karlsson, and P.~A. Andrekson, ``Phase-coherent lightwave communications
	with frequency combs,'' \emph{Nat. Commun.}, vol.~11, no.~1, p. 201, Jan.
	2020.
	
	\bibitem{8576586}
	A.~F. {Alfredsson}, E.~{Agrell}, H.~{Wymeersch}, B.~J. {Puttnam},
	G.~{Rademacher}, R.~S. {Luís}, and M.~{Karlsson}, ``Pilot-aided
	joint-channel carrier-phase estimation in space-division multiplexed
	multicore fiber transmission,'' \emph{J. Lightw. Technol.}, vol.~37, no.~4,
	pp. 1133--1142, Feb. 2019.
	
	\bibitem{7328945}
	R.~S. Lu\'{i}s, B.~J. Puttnam, J.-M. {Delgado Mendinueta}, Y.~Awaji, and
	N.~Wada, ``Impact of spatial channel skew on the performance of
	spatial-division multiplexed self-homodyne transmission systems,'' in
	\emph{Proc. International Conf. Photon. Switching}, Sep. 2015, pp. 37--39.
	
	\bibitem{1312647}
	O.~{Simeone} and U.~{Spagnolini}, ``Adaptive pilot pattern for {OFDM}
	systems,'' in \emph{Proc. {IEEE} International Conf. Commun.}, vol.~2, Jun.
	2004, pp. 978--982.
	
	\bibitem{6585733}
	M.~{Simko}, P.~S.~R. {Diniz}, Q.~{Wang}, and M.~{Rupp}, ``Adaptive pilot-symbol
	patterns for {MIMO OFDM} systems,'' \emph{{IEEE} Trans. Wireless Commun.},
	vol.~12, no.~9, pp. 4705--4715, Sep. 2013.
	
	\bibitem{5419953}
	P.~{Fertl} and G.~{Matz}, ``Channel estimation in wireless {OFDM} systems with
	irregular pilot distribution,'' \emph{{IEEE} Trans. Signal Process.},
	vol.~58, no.~6, pp. 3180--3194, Jun. 2010.
	
	\bibitem{5946248}
	M.~D. {Larsen}, G.~{Seco-Granados}, and A.~L. {Swindlehurst}, ``Pilot
	optimization for time-delay and channel estimation in {OFDM} systems,'' in
	\emph{Proc. International Conf. Acoustics, Speech, Signal Process.}, May
	2011, pp. 3564--3567.
	
	\bibitem{8108382}
	Y.~{Zhang}, J.~{Liu}, S.~{Feng}, and P.~{Zhang}, ``Pilot design for phase noise
	mitigation in millimeter wave {MIMO-OFDM} systems,'' in \emph{Proc. Vehicular
		Technol. Conf.}, Jun. 2017.
	
	\bibitem{6236002}
	J.~{Lee}, Y.~{Ha}, B.~{Shin}, S.~{Kim}, B.~{Kim}, and W.~{Chung}, ``Novel pilot
	scheme for transmitter {IQ} mismatch compensation in {CO-OFDM} system,''
	\emph{{IEEE} Photon. Technol. Lett.}, vol.~24, no.~17, pp. 1543--1545, Sep.
	2012.
	
	\bibitem{8052163}
	X.~{Ma}, H.~{Zhang}, X.~{Yao}, and D.~{Peng}, ``Pilot-based phase noise, {IQ}
	mismatch, and channel distortion estimation for {PDM CO-OFDM} system,''
	\emph{{IEEE} Photon. Technol. Lett.}, vol.~29, no.~22, pp. 1947--1950, Nov.
	2017.
	
	\bibitem{ZHANG201814}
	H.~Zhang, X.~Ma, and P.~Li, ``A novel transmitter {IQ} imbalance and phase
	noise suppression method utilizing pilots in {PDM CO-OFDM} system,''
	\emph{Opt. Commun.}, vol. 413, pp. 14--18, Apr. 2018.
	
	\bibitem{4562696}
	J.~{Bhatti} and M.~{Moeneclaey}, ``Influence of pilot symbol configuration on
	data-aided phase noise estimation from a {DCT} basis expansion,'' in
	\emph{Proc. International Netw. and Commun. Conf.}, May 2008, pp. 79--84.
	
	\bibitem{alfredsson2017ecoc}
	A.~F. Alfredsson, E.~Agrell, H.~Wymeersch, and M.~Karlsson, ``Pilot
	distributions for phase tracking in space-division multiplexed systems,'' in
	\emph{Proc. European Conf. Opt. Commun.}, Sep. 2017, p. P1.SC3.48.
	
	\bibitem{johnson:sampling:2012}
	S.~K. Johnson, \emph{Sampling}, 3rd~ed.\hskip 1em plus 0.5em minus 0.4em\relax
	Hoboken, NJ, USA: John Wiley \& Sons, 2012.
	
	\bibitem{Alvarado:18}
	A.~Alvarado, T.~Fehenberger, B.~Chen, and F.~M.~J. Willems, ``Achievable
	information rates for fiber optics: Applications and computations,'' \emph{J.
		Lightw. Technol.}, vol.~36, no.~2, pp. 424--439, Jan. 2018.
	
	\bibitem{1021039}
	K.~Cho and D.~Yoon, ``On the general {BER} expression of one- and
	two-dimensional amplitude modulations,'' \emph{{IEEE} Trans. Commun.},
	vol.~50, no.~7, pp. 1074--1080, Jul. 2002.
	
\end{thebibliography}
\end{document}